\documentstyle[prl,aps]{revtex}
\begin{document}
\input{psfig.sty}
\def \gam {\frac{ N_f N_cg^2_{\pi q\bar q}}{8\pi} }
\def \gamm {N_f N_cg^2_{\pi q\bar q}/(8\pi) }
\def \be {\begin{equation}}
\def \ba {\begin{eqnarray}}
\def \ee {\end{equation}}
\def \ea {\end{eqnarray}}
\def \gap {{\rm gap}}
\def \gapp {{\rm \overline{gap}}}
\def \gappp {{\rm \overline{\overline{gap}}}}
\def \im {{\rm Im}}
\def \re {{\rm Re}}
\def \Tr {{\rm Tr}}
\def \P {$0^{-+}$}
\def \S {$0^{++}$}
\def \uu {$u\bar u+d\bar d$}
\def \ss {$s\bar s$}
\title{\hfill To appear in Phys. Lett. B, hep-ph/9712479 \\ \vskip .5cm  
Spontaneous Generation of Pseudoscalar Mass 
in the U(3)$\times$U(3) Linear Sigma Model }
\author{Nils A. T\"ornqvist}
\address{Physics Department, 
POB 9, FIN--00014, University of Helsinki, Finland\footnote {e-mail address:
Nils.Tornqvist@Helsinki.Fi}}
\date{Dec. 20th 1997, revised  March 15th 1998}                 
\maketitle
\begin{abstract}
A novel, nonperturbative, way to generate chiral symmetry breaking within the 
 linear sigma model for 3 flavours with an interaction term $\lambda 
{\rm Tr} [\Sigma\Sigma^\dagger\Sigma\Sigma^\dagger]$ is discussed. After 
spontaneous chiral symmetry
breaking in the vacuum at the tree level the scalar nonet obtains mass,
while the pseudoscalars are massless. Then, including quantum loops in
a nonperturbative, self-consistent way chiral symmetry is broken 
by nonplanar graphs in a second
step, and also the pseudoscalars become massive. By interpreting the basic 
symmetry to be a discrete permutation symmetry, in accord with superselection
rules, no additional Goldstone bosons are expected. 
\\
\vskip 0.05cm 
\noindent Pacs numbers:12.39.Ki, 11.30.Hv, 11.30.Qc, 12.15.Ff
\vskip 0.20cm 
\end{abstract}

{\it 1. Introduction and the linear sigma model.}
Today the detailed experimental data on the light scalar and pseudoscalar mass
spectrum and mixings defy any simple phenomenological explanation. It seems
obvious that these mesons require a much better
understanding of the nonperturbative and nonlinear aspects of QCD at low
energies, than we have today. When hopefully in the near future these mesons 
are understood, we most probably  have a much better understanding also of the
confinement mechanism. This paper is an attempt to bridge this gap, 
and to understand hadron mass generation  in general.
 
 I shall first argue that a 
good candidate for an effective meson theory at low energies, when the
gluonic degrees of freedom are integrated out, is the
generalization of the well known linear sigma model\cite{gm,lee} to $U(3)\times
U(3)$ including one scalar and one pseudoscalar nonet.   
I restrict myself generally to 3 light flavours ($N_f=3$), although sometimes I
keep the $N_f$ in the formulas for clarity. Consider 
the basic $U(3) \times U(3)$  symmetric, classical Lagrangian,
which has same flavour and chiral symmetries as QCD:

\be  {\cal L}=
\frac 1 2 \Tr [\partial_\mu\Sigma \partial_\mu\Sigma^\dagger]
+\frac 1 2 m^2\Tr [\Sigma \Sigma^\dagger] -\lambda \Tr[\Sigma\Sigma^\dagger
\Sigma\Sigma^\dagger]  -\lambda' (\Tr[\Sigma\Sigma^\dagger])^2 
+{\cal L}^{SB} \ .
\label{lag}
\ee

Here $\Sigma= \sum_{a=0}^8(s_a+ip_a)\lambda_a/\sqrt 2$ are $3\times 3$
matrices, $s_a$ and $p_a$ stand for the \S\  and \P\  nonets and
$\lambda_a$ are the Gell-Mann matrices, normalized as $\Tr[\lambda_a\lambda_b]=
2\delta_{ab}$, and where for the singlet 
$\lambda_0 = (2/3)^{1/2} {\bf 1}$ is 
included. Note that each meson from the start has a definite $SU(3)$ symmetry
content, which in the quark model means that it has a definite $q\bar q$
content. Thus the potential terms in Eq.~(1) can be given
 a conventional quark line structure shown in Fig.~1.

\begin{figure}[h]
\centerline{
\protect
\hbox{
\psfig{file=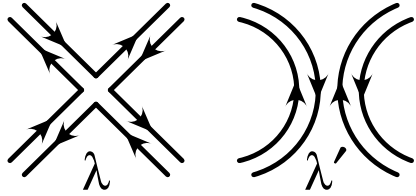,width= .3000\textwidth,angle=0}}}
\caption{Graphical quark line 
representation of the $\lambda$ and $\lambda '$ terms
of Eq.~(1). Note the disconnected quark line nature of the $\lambda '$ term.}
\end{figure}

 Apart from the symmetry breaking term ${\cal L}^{SB}$, 
Eq.~(1) is clearly invariant under $\Sigma \to U_L\Sigma U_R^\dagger$ of
$U(3)\times U(3)$.
In Eq.~(1)  I have contrary to the usual convention defined the sign of $m^2$
such that the naive physical squared mass would be $-m^2$, and the instability
thus occurs when $m^2>0$.

If  $-m^2<0$ and $\lambda>0, \lambda'>0 $
 the potential in Eq.~(1)
(being of the form of a "Mexican hat") gives rise to an instability with vacuum
condensate $<s_0>=(2/3)^{1/2}f_\pi$. Let $\lambda' =0$.  Then shifting
as usual the scalar field, $\Sigma \rightarrow \Sigma + f_\pi{\bf 1} $, 
one finds
$f_\pi^2 =m^2/(4\lambda)$ (cf. Fig.~2). 
Furthermore, the squared mass ($-m^2$) of the \S\ nonet
is replaced nonperturbatively  by $-m^2+12\lambda f_\pi^2=2m^2$, while 
the \P\ nonet  becomes massless,
$-m^2+4\lambda f_\pi^2 =0$. The symmetry of the spectrum is broken down to 
$SU(3)\times U_A(1)$. If we had included the $\lambda'$ term 
instead of the $\lambda $ term, then only the scalar singlet would aquire
mass while the remaining 17 states would be massless. 
Then more symmetry
remains in the spectrum; O(18) symmetry is broken down to O(17).
 More generally with both $\lambda$ and $\lambda'$ present the
scalar octet squared mass is $2m^2\lambda/(\lambda+3\lambda')$, while
the singlet squared mass is $2m^2$, and the 9 pseudoscalars are  massless.  

\begin{figure}[h]
\centerline{
\protect
\hbox{
\psfig{file=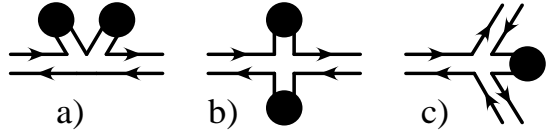,width= .400\textwidth,angle=0}}}
\caption{ After shifting the field $\Sigma  \to \Sigma +f_\pi $ 
(denoted by the blob)
 the $\lambda $ term in Fig.~1 generates the mass terms in  a and b, 
$[4\Tr\Sigma\Sigma^\dagger +\Tr(\Sigma\Sigma+h.c.)]\lambda f_\pi^2$, 
and the trilinear couplings, 
$4\lambda f_\pi \Tr [\Sigma\Sigma\Sigma^\dagger + h.c.]/2$
shown in c, which are flavour symmetric and obey the  OZI rule. }
\end{figure}

In addition, very importantly, after  shifting the scalar singlet field,  
the $\lambda$  term (keeping $\lambda'=0$) generates trilinear $spp$ and $sss$
couplings of the form $g\Tr [\Sigma\Sigma\Sigma^\dagger +h.c]/2$,
(cf. Fig.~2c), where g is  $g=4\lambda f_\pi$. With the flavour indices 
written out  explicitly one has (not including the combinatoric factors) 
\be
g_{abc}=g\Tr [\lambda_a \lambda_b \lambda_c
+\lambda_a\lambda_c\lambda_b ]/(2\sqrt 8) \ .
\ee
 These couplings   obey
the simply connected, Okubo-Zweig-Iizuka  (OZI) allowed,
quark line rules with flavour symmetry
exact. One has SU(3)$_f$ predictions relating different couplings constants.
Denoting by $\sigma$ the \uu\ scalar, and by $\sigma_s$ the \ss\ scalar one has
e.g. $g^2 =2g^2_{\sigma\pi^0\pi^0}=  g^2_{\sigma\pi^+\pi^-} 
=2g^2_{\sigma K\bar K}=
\frac 4 3 g^2_{K^*_0K\pi}= 2g^2_{a_0K\bar K} =  g^2_{\sigma_sK\bar K}$,
and  $g_{\sigma_s\pi\pi}=0$ etc. Here we summed over charge states except for
the $\sigma\pi\pi$ couplings, since conventionally the $\sigma\pi\pi$ coupling
is $g_{\sigma\pi^0\pi^0}$. If one includes also the $\lambda'$ term
of Eq.~(1) then only the couplings involving the $\sigma$ and $\sigma_s$ 
states would be altered, which would violate the OZI rule at the tree level.

Conventionally one includes  into $\cal{L}$ terms which break the
symmetries:
\be
{\cal L}^{SB}=\epsilon_0 s_0 + \epsilon_8 s_8 +c \cdot [ \det \Sigma +
\det \Sigma^\dagger] \ .
\ee
 
Here $\epsilon_0$ gives, by hand, the pseudoscalar nonet a common mass,
while  $\epsilon_8$
breaks explicitly the remaining $SU(3)$ \  down to isospin.
These terms are related to quark masses in QCD.
Because of the quantum effects in QCD involving the gluon anomaly the 
$U_A(1)$ symmetry of Eq.~(1) is broken. This is represented by 
the $c$ term in Eq.~(3), which gives the $\eta_1$ an extra  mass\cite{hooft}. 
For most of
the discussion below we shall neglect the $\epsilon_8$ and $c $ terms,
 except in the discussion
of the fit to the scalar mesons below, 
where they enter through the \P\  masses.

{\it 2. The {\rm U3$\times$U3} sigma model and  scalar meson data.}
In this section I include some results of  phenomenology for two reasons: 
(i) I want to  emphasizes the fact that the U3$\times$U3 model discussed above 
is in fact phenomenologically a very successful model for the scalar mesons, 
and (ii) I need the parameters determined here  
in order to estimate the pseudoscalar mass in Eq.~(11) below.   

The  flavour symmetric OZI rule obeying  
couplings of Eq.~(2) together with 
a near-degenerate bare scalar nonet mass
were, in fact, the starting point of our recent analysis 
of the scalar $q\bar q$
nonet \cite{NAT}. In particular it was crucial that after determining the
overall coupling $g$ from a fit to data on  
 $K^*_0\to K\pi$ and $a_0\to\pi\eta$ one essentially predicted the $\pi\pi$ 
I=0 S-wave phase shift (Fig.~3). 
This shows that the above relations relating  the
bare $\sigma$ and $\sigma_s$ couplings to the same overall $g$ as those of
$K^*_0K\pi$ and $a_0\pi\eta$ must be approximately
satisfied experimentally. I.e., one cannot tolerate a 
very big bare $\lambda'$ coupling
in Eq.~(1) since then these relations would be  destroyed. 
Another argument for
that $\lambda'$ must be small is that then the bare scalar singlet
 state would have a very different mass from the other nonet members, 
not needed in the fit.
 After the unitarization the scalars aquired finite widths
and were strongly shifted in mass by the different couplings to the \P \P\
thresholds. The \P\  masses in the thresholds were given their experimental
values (i.e. we included effectively ${\cal L}^{SB}$ for the \P\ states), 
and consequently the main source of flavour symmetry breaking in the
output physical mass spectrum was generated by the vastly different positions
of these thresholds. 
E.g. the large experimental splitting between the $a_0(980)$
and $K^*_0(1430)$ masses came from the large breaking in the sum of
loops for the $K^*_0$ ($K^*_0\to K\pi , K\eta , K\eta' \to K^*_0$) compared to
those for $a_0$ ($a_0\to \pi\eta , K\bar K , \pi \eta' \to a_0$),
although in the strict $SU(3)_f$
limit these thresholds would lie on top of each others and would then
together give the same mass shift to the two resonances.

There were only 6 parameters in \cite{NAT} out of which two parametrized the
bare scalar spectrum (1.42 GeV for the $u\bar u$, $d\bar d$, $u\bar d$,
$d\bar u$, and an extra 0.1 GeV when an $s$ quark replaces a $u$ or $d$
quark). 
The overall coupling was parametrized by $\gamma=1.14$, and $k_0=0.56$ GeV/c
was the cut off parameter. 
Now, the $\gamma$ parameter can be related to $\lambda$
of Eq.~(1) through $\lambda=4\pi\gamma^2 =16$, by comparing
the $\sigma \pi\pi$ coupling of the two schemes. [One has
$g^2_{\sigma\pi^0\pi^0}/(4\pi)$ $ = \gamma^2 m^2_{\sigma}= \lambda
m^2_{\sigma}/(4\pi)$. The latter equation follows from $g_{\sigma\pi^0\pi^0}
=g/\sqrt 2=4f_\pi \lambda$  and $f_\pi^2=m^2/(4\lambda )=m_\sigma^2/(8\lambda)$ given
above.]

\begin{figure}[h]
\centerline{
\protect
\hbox{
\psfig{file=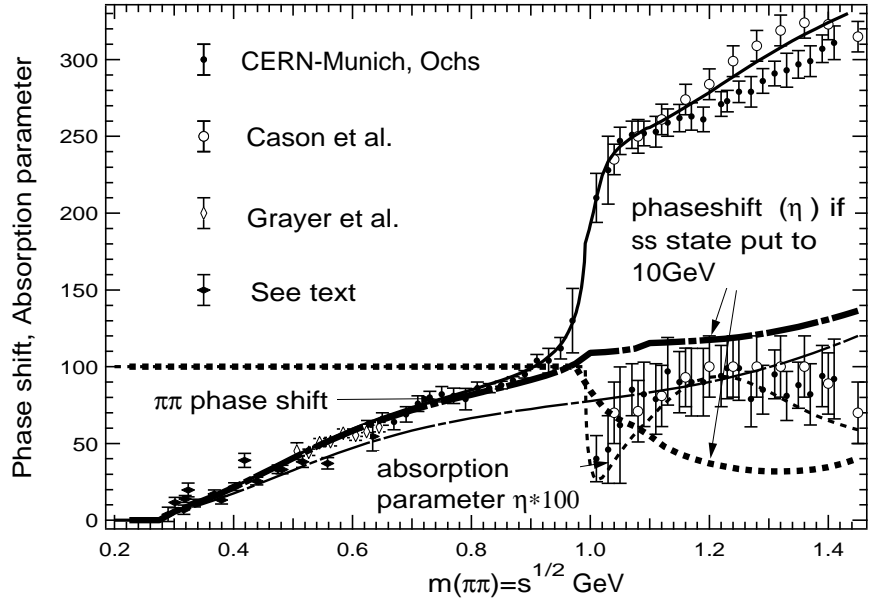 ,width=.70\textwidth,angle=0}}}
\caption{The full line curve shows the 
I=0, S-wave phase shifts for $\pi\pi\rightarrow \pi\pi$, as predicted by 
the  model [3]. If one lets the $s\bar s$ meson mass
become very large (10 GeV) then the rapid rise at 1 GeV
due to the $f_0(980)$ vanishes. This shows that $f_0(980)$ is the unitarized
$s\bar s$ state. The remaining phase shift (dot-dashed thick line) is
then entirely due to the $u\bar u +d\bar d$ channel and one sees that what
remains can be understood as a very broad resonance. This is  the "$\sigma$", 
whose 90$^\circ$ mass value is at 880 MeV, while the pole is at 470-i250 MeV.
(The thin dot-dashed curve shows the phase shift if also the $K\bar K$ and
$\eta\eta$ thresholds are given large values.) }
\end{figure}                                    
                                                
Using the conventional value for $f_\pi =93$ MeV 
and $\lambda=16$ one predicts from (1) at the tree-level
 that $m_\sigma = f_\pi (8\lambda )^{1/2} $=1060 MeV as an average mass of the 
$u\bar u, \ d\bar d,\ d\bar u $ and $u\bar d $ states. Now although
strictly speaking there is no exact one-one correspondence between the
 model of Eq.~(1) at the tree level, before unitarization, and the
unitarised model of \cite{NAT}, it is remarkable that this prediction
is close to the average mass of the 
$a_0(980)$ and $\sigma$ resonances  found in the fit.
 I.e. if we had used $f_\pi=93$ MeV to determine the energy  scale,
we could have eliminated one of the 6 parameters.
  
An important point observed in the second paper of Ref. \cite{NAT} was that
the model requires the existence of the light and broad $\sigma$ resonance 
pole. This is explained in more detail in Fig.~3.
Recently there has appeared 
several new  papers \cite{ishi}, 
which through different analyses and models  support this same  
conclusion, i.e.
that the light and broad sigma, which has been controversial for so long,
really exists, and is here to stay \cite{PDG}. 

The important conclusion of this phenomenological section is that the model of 
\cite{NAT} can be
interpreted as an effective field theory  given by  Eqs.~(1) and (3)
with $\lambda=16$ and $\lambda'\approx 0$. (For other determinations of
$\lambda$ and $\lambda '$ where $\lambda'$ generally is not small see 
\cite{pisarski}-\cite{chan}).  The absence of the $\lambda'$
term at the tree level means that the OZI rule holds exactly at the tree level.
Of course the unitarization procedure can be improved upon,
including u- and t-channel singularities \cite{isgur}, $s\to ss\to s$ loops,
higher order diagrams etc., but I am confident that the dominant effects were
already included phenomenologically for the scalar states. 

{\it 3. Spontaneous generation of pseudoscalar mass.}
Let us now consider loops and renormalization.  As is 
well known the linear sigma model is
renormalizable, as  first shown by Lee\cite{lee}. 
Crater\cite{crater} discussed the
renormalization of the $U(3)\times U(3)$ symmetric model, Eq.~(1), 
before spontaneous breaking and without
${\cal L}^{SB}$, and showed that the $\lambda$ term is not alone
renormalizable but requires through loops
the presence of the $\lambda'$ coupling. This is not
inconsistent with the result above that the bare $\lambda'$ term must be small.
Paterson\cite{paterson} has showed that the Coleman-Weinberg\cite{cw}
 mechanism occurs when $\lambda\neq 0$ 
 i.e, that the symmetry is spontaneously broken also in the case when the 
bare mass term $m $  is assumed  to vanish. 
Thus the $\lambda$ term generally requires, 
after radiative corrections, the presence
of both a nonvanishing 
mass term and a small $\lambda'$ term in the 
renormalized Lagrangian.
Furthermore, Chan and Haymaker\cite{chan} 
have discussed the renormalizability of  the full Lagrangian (1), with 
the symmetry breaking term (3),  and  $<s_0>\neq 0$ 
present from the start. 

Consider the lowest order loops generated by the Lagrangian (1) after shifting
the scalar field, Figs.~4a,b and 5a,b. 
In these one loop integrals one  needs the two familiar
functions $A(m^2)$ and $B(m_1^2,m^2_2)$, given below with a three momentum 
cut off. The simple function $A(m^2)$ is the same one-loop function which appears in
most gap equations, c.f.\cite{nambu,NJL}.                 
\ba A (m^2) &=& \int \frac
{i d^4k}{(2\pi)^4}[k^2-m^2+i\epsilon]^{-1} = \frac 1 {32 \pi^2}
\int^{4(\Lambda^2+m^2)}_{4m^2} (1-4m^2/s)^{\frac 1 2}ds  \cr 
&=& \frac
{\Lambda^2} {8 \pi^2} \Bigl [(1+\frac{m^2}{\Lambda^2})^{\frac 1 2}  -\frac
{m^2} { \Lambda^2} \ln [ \frac \Lambda m  (1+(1+\frac{m^2}{\Lambda^2})^{\frac 1
2} ) ]  \Bigr ]  \ . \label{A} 
\ea
\ba
B(m_1^2,m^2_2) &=& \int \frac {id^4k}{(2\pi)^4}[(k^2-m_1^2+i\epsilon)
(k^2-m_2^2+i\epsilon) ]^{-1}  
=(m_1^2-m_2^2)^{-1} [A(m_1^2)-A(m_2^2)] \ .\label{B}
\ea                                                   
Denote the degenerate \P\ octet masses (i.e. $\pi,K,\eta_8$) by $m_{p_8}$
 and the 
degenerate \S\ octet masses ($a_0,K^*_0,\sigma_8$) by $m_{s_8}$, 
while the singlet mases  are
$m_{p_0}$ and $m_{s_0}$. When one can neglect the octet singlet splitting
the two nonet masses  are denoted $m_{s_9}$ and $m_{p_9}$.

\begin{figure}[h]
\centerline{
\protect
\hbox{      
\psfig{file=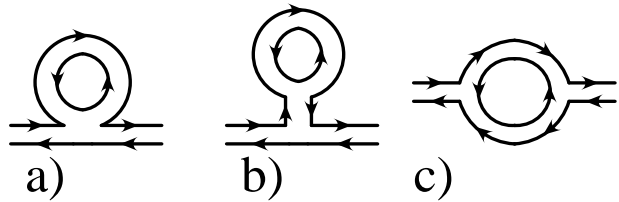,width= .4500\textwidth,angle=0}}}
\caption{The planar tadpole  diagram (a) generated from the $\lambda$ term, 
the planar tadpole diagram (b) with an intermediate scalar singlet
and the planar loop diagram (c) generated from the trilinear couplings in
Fig.~2c. When flavour symmetry is unbroken these diagrams 
contribute equally to each member of a nonet, 
and the internal closed loop simply gives a factor
$N_f$. Thus they have the same  structure as the mass terms in
the tree level Lagrangean and can be included in it as renormalization terms.
 } 
\end{figure}

\begin{figure}[h]
\centerline{
\protect
\hbox{
\psfig{file=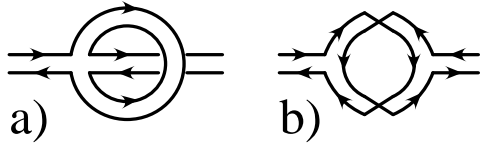,width= 0.350\textwidth,angle=0 }}}
\caption{The nonplanar, OZI-rule violating  
one loop graphs, which contribute only to the singlet channels. 
These are nonvanishing when $<s_0>\neq 0$ and the scalars are massive and
nondegenerate with the pseudoscalars.
 For the \P\ singlet the sum of the one-loop digrams cancel 
(Eq.~7), but for the scalar singlet they add (Eq.~8)
renormalizing the scalar singlet quadratic term 
differently from the rest. }
\end{figure}

There are two main classes of diagrams: the "planar"  diagrams of Fig.~4
and the disconnected OZI rule violating "nonplanar"
 diagrams of Fig.~5. As long as
$SU(3)_f$ remains unbroken one can sum over flavour in the planar
graphs giving simply a factor $N_f$.
Therefore these planar diagrams at most only  renormalize the
masses of the two nonets. For the pseudoscalar  nonet one gets, 
(at $s=p^2=0$)
\ba
\Delta m^2_{p9}(planar)&=& N_f[(8\lambda -\frac{6g^2}{m^2_{s0}}) 
A(m^2_{s_9}) +(8\lambda -\frac{2g^2}{m^2_{s0}}) 
A(m^2_{p_9})+2g^2 B(m^2_{s_9},m^2_{p_9})]= \cr
&=&\lambda N_f[(8-12+4)A(m^2_{s9})+(8-4-4)A(m^2_{p9})]\to 0 \ ,  \label{delm}
\ea
The loops generated directly from the $\lambda$ term Fig.~4a and 5a have a 
combinatoric factor of 12 out of which 8 are planar and 4 nonplanar.
Therefore the two numbers 8 in Eq.~(\ref{delm}).
 On the other hand the tadpole loops of Fig. 4b contribute the 
terms with the numbers $-12$ and $-4$ (when one uses the relation 
$2g^2/m^2_{s0}=4\lambda$). Finally the planar loops Fig.~4c, which give the 
$2g^2B$ term, contribute
with the numbers $-4$ and $+4$, when one furthermore 
uses the relation (\ref{B}). Similar 
cancellations whithin the standard sigma model including $\pi$ and $\sigma$
only have been studied by Bramon et al. \cite{ishi}, who also showed that 
if one adds quarks to the theory, and considers diagrams 
like Fig.~4b and 4c, but whithout the inner closed 
loop, also  these diagrams cancel each other.

For the \S\ nonet one gets in a similar way a common mass shift to the 
whole nonet, which can be included into the bare nonet mass.
  One can consider the loop diagrams of Fig.¨4c as the driving 
terms of the instability,
which contribute to the negative curvature of the effective potential 
near the origin, and to the ``wrong'' sign $m^2$ term in Eq.¨(1). 
The tadpole terms 
then corresponds to how the vacuum responds, i.e. to the  terms obtained 
through the vacuum condensate.  
Thus if we would restrict ourselves to including 
planar diagrams only, we would have a similar situation as at the tree
level, with massless pseudoscalars and a massive scalar nonet. 

However, there are of course also nonplanar diagrams, Fig.~5,
which contribute only to the flavour singlet channels. Here the situation is
more interesting (although in the large $N_c$ limit of QCD
these would vanish).  
 These diagrams give an extra contribution to the 
scalar singlet channel or the vacuum channel, which 
determines the vacuum condensate. 
There is an important minus sign  whenever the nonet in the loop of Fig.~5a 
has the opposite parity than the external meson. 
This sign change can be seen from the
negative sign of the last term in the expansion  (see e.g.\cite{gasi})
$\Tr [\Sigma \Sigma^\dagger \Sigma \Sigma^\dagger ] 
=\Tr [S^4] +\Tr[P^4] +4\Tr [S^2P^2]-2\Tr[PSPS]$,
where $\Sigma =S+iP$, i.e., $S=\sum_a
s_a\lambda_a/\sqrt 2$ and $P=\sum_a p_a\lambda_a/\sqrt 2$.

One  can sum over scalar and pseudoscalar nonets in the loops of 
Fig.~5, and find for the \P\ singlet:
\be
\Delta m^2_{p_0}(nonplanar) =N_f \bigl [-4\lambda  A
(m^2_{s_9})+4\lambda A(m^2_{p_9}) +2g^2B(m^2_{p_9}, m^2_{s_9})\bigr
]\to  0\label{dm0} \ .
\ee 
This sum vanishes exactly, 
when (i) the pseudoscalar masses vanish and (ii) when one neglects
the small second order splitting between $s_8$ and $s_1$,
 because of the relation $g^2=2\lambda
m^2_{s_9}$ (which, by the way, is 
the same relation which gives the Adler zeroes in \P\P\ scattering).
Then the contribution from Fig.~5a cancels that from Fig.~5b 
as seen from the relation between $A$ and $B$ in Eq.~(\ref{B}).

But, for the scalar singlet channel the signs of the two tadpole terms are
opposite  to that in Eq.~(\ref{dm0})
(again because of the important minus sign mentioned above) and
one finds a nonvanishing result: 
\ba
-\Delta\bar m^2_0 &=& \Delta m^2_{s_0}(nonplanar) \cr 
&=&N_f \bigl [4\lambda  A
(m^2_{s_9})-4\lambda A(m^2_{p_9}) +
g^2B(m^2_{p_9},m^2_{p_9})+9g^2B(m^2_{s_9},m^2_{s_9}) 
\bigr]\neq 0 \ .  \label{dms0}
\ea 

If one would evaluate  this quantity  using in the original
Lagrangian, before shifting the scalar singlet field, 
i.e. when  still $m_{s_9}=
m_{p_9}$ and $g=0$, one would get zero also for this quantity.
But, once the scalars and pseudoscalars are split in mass by the
chiral symmetry breaking in the vacuum the loops in Fig.~5 are not anymore 
vanishing,  and contribute to making $\Delta\bar m^2_0 \neq 0$. 
Now I argue this does not only shift the scalars singlet mass slightly 
down from the scalar octet,
but more importantly, because of self-consistency it also 
increases the instability in the scalar channel and thus
  also contributes to the shape of the potential, in a
``second step of chiral symmetry breaking'': 

{\it The renormalized 
curvature of the potential in the scalar singlet direction
will be bigger than in the other directions, i.e., "the Mexican hat will be
warped" by the extra quadratic term $-\frac 1 2 \Delta \bar m^2_0s_0^2$. }

The nonvanishing of  $-\Delta\bar  m^2_0$ 
in Eq.~(8) is crucial in the following. 
It  has the right negative sign
making  the quadratic term for 
the scalar singlet  more negative
than  the corresponding term for the pseudoscalar nonet, 
and  also more  negative than the quadratic term for the scalar octet. 
This guarantees that the minimum of the renormalized potential will
be in the direction of the scalar singlet, and that only chiral symmetry, not
flavour nor parity, is violated in the solution.

Including this term into  
the stability condition that the linear term involving the scalar singlet
should vanish one finds for the renormalized  $f_\pi$. 
\be
f_\pi^2 = (\bar m^2 + \Delta \bar  m_0^2)/(4\lambda ) \ ,
\ee
where  all quantities
$  m^2,\ \Delta \bar m^2_0$ and $\lambda$ are 
defined such that
they normally are positive, when spontaneos symmetry breaking occurs. 

Summing the different contribution to the four  masses one finds 

\be \begin{array}{rlllll}
m^2_{s_0} &=  - m^2 -\Delta \bar m^2_0+ 12\lambda f_\pi^2 
          &=2\Delta\bar  m^2_0 +2 m^2
          &=8\lambda f_\pi^2            \label{ms0} \ ,\\
m^2_{s_8} &= - m^2 + 12\lambda f_\pi^2
          &=3\Delta\bar  m^2_0  +2 m^2 
          &=8\lambda f_\pi^2 +\Delta\bar  m^2_0        \ ,    \\
m^2_{p_0}\approx m^2_{p_8} &= - m^2 + 4\lambda f_\pi^2 
          &= \  \Delta \bar m^2_0     \ .   
\end{array}
\ee

Once the \S\ states are split from the \P\ states through the first step of
chiral symmetry breaking, then as a second step the nonplanar   loops 
renormalize the potential with an extra quadratic term in the scalar 
singlet direction, $-\frac 1 2  \Delta m^2_0 s_0^2$.
\begin{figure}[h]
\centerline{
\protect
\hbox{
\psfig{file=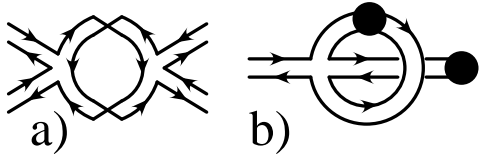,width= .3500\textwidth,angle=0}}}
\caption{The one loop diagram (a) generated from the $\lambda$ term,
gives a contribution of the same disconnected flavour structure as the
$\lambda '$  term of Fig.~1b.  Thus renormalization
requires the presence of the $\lambda'$ term in Eq.~(1).
Fig.~6b  shows the tadpole graph
which contributes to the scalar singlet channel.
The blob on the internal line indicates that the vacuum insertions of Fig.~2a,b
and loops of Fig.~5 should iteratively be included into the internal lines.
This diagram is then  nonvanishing if $<s_0>\neq 0$, 
when the scalars are massive and
consequently the  \P\ and the \S\ states in the loop do not cancel.
It has near the minimum the
same flavour structure as the $\epsilon_0 s_0$ term in Eq.~(3). 
Thus renormalization requires its presence
in and it gives the pseudoscalars mass. }
\end{figure}

It is of course very well known that renormalization deforms  
the effective potential from that of the tree level. Also, the fact that
renormalization often requires the presence of new terms is well known.
An example of the latter is provided by the fact 
that the $\lambda$ term of Eq.~(1) requires the presence of a (small)
$\lambda'$ term \cite{crater}, which we most easily  can  see
graphically from Fig.~6a. The one 
loop correction generated from the $\lambda$ 
term shown in Fig.~6a has the same disconnected flavour structure as the
$\lambda '$ term of Eq.~(1) and Fig.~1b. The unconventional new result 
presented here is that quantum effects, through the 
self-consistency condition, 
can also, nonperturbatively, generate new
terms which violate the original symmetries of the tree Lagrangian. 

 Even this result is not quite new, the breaking of the $U_A(1)$
symmetry by the anomaly is another example of symmetry breaking through
quantum effects. And in fact, the new mechanism also breaks the $U_A(1)$
symmetry albeit in a new and simpler way. 
The main  difference  is that not only the
$\eta_1$, but the whole \P\  nonet aquires mass, and that the effective 
potential obtains a quadratic symmetry breaking term, which warps the
potential. Near the minimum the quadratic warping
can be replaced by a conventional linear $\epsilon_0 s_0$ term, as in Eq.~(3),
represented graphically by the nonvanishing diagram in Fig.~6b.
Because of this  the \P\  nonet 
obtains a mass $\Delta\bar  m_0$.
 Near the minimum this has the same effect as
the conventional explicit symmetry breaking term $\epsilon_0 s_0$
with  $\epsilon_0 \approx \Delta\bar m_0^2 (2/3)^{1/2} f_\pi$.

{\it 4. Predicted pseudoscalar mass.} What is the magnitude of the predicted 
pseudoscalar mass? It is clear that it
really should depend only on the dimensionless coupling
$\lambda$\ evaluated at the appropriate scale, and the scalar mass 
(or $f_\pi$). In a forthcoming publication I shall make a more detailed 
instability calculation. Here,  as a rough approximation choose 
the same parameters as in the fit to the scalar nonet: 
$\lambda (\approx 1 $ GeV) =16, $\Lambda =560$ MeV/c, 
for the average nonet mass $m_{s_9} \approx 1 $ GeV and for the 
input average pseudoscalar mass a range between 0 and 500 MeV.
One finds from Eqs.~(8) and Eq.~(10)
(neglecting the $g^2B$ terms which would  increase 
the predicted \P\ mass somewhat)  for $N_f=3$
\be m_{p_8} \approx 
\bigl [4N_f\lambda [A(m^2_{p_8})-A(m^2_{s_8})]\bigr ]^{1/2} 
=  450\pm 200 {\rm MeV}. \ee
Using for the average \P\ input mass 450 MeV
one gets also 450 MeV for the output.
This can be compared with the average experimental pseudoscalar
octet mass of 368 MeV.
This estimate is probably fortuitous, but already the fact that one gets the
right order of magnitude is highly nontrivial, and shows that my
mechanism can predict reasonable  \P\ masses. 
Certainly, the comparison with experiment 
can be improved upon by a more detailed calculation, and by including 
$SU(3)_f$ breaking, vector mesons, the running of the coupling  
$\lambda (\mu )$
etc. Qualitatively one expects that the running of coupling constant,
 in analogy with $\phi^4$ theory, decreases $\lambda$ as one moves
from the 1 GeV region where it was determined down to the pseudoscalar 
masses. This would also reduce the predicted
\P\ masses. But the calculation of the $\beta$ function for the
present model is a complicated matter indeed, especially as one would have to
keep the detailed analytical threshold behaviour for  the large number of 
thresholds involved.

{\it 5.  The self-consistency condition.} 
The essential  condition, which I have imposed                 
is that the same  physical masses should be used for 
the "input" masses in the loops
on the r.h.s. of the equations as obtained for the 
"output" physical masses on the l.h.s of Eqs.~(\ref{ms0}). 
This is as any self-consistency condition "circular" in the sense 
that one way  of solving it is by  iteration, 
inserting the output masses into the input masses. This generates 
diagrams with loops and vacuum insertions within loops 
{\it ad infinitum}.

With this condition the potential is deformed by quantum 
corrections\footnote{One may
look at this two-step breaking of chiral symmetry using the well known Mexican
hat analogy. Instead of having the sombrero on a table, hang it on a peg at its
middle. Then as the ball falls from the labile position at the top of the
hat into the brim (first step) it tilts, or warps, the hat in a
 second step of symmetry breaking. (In the actual model only the warp occurs).
 A unique minimum is created giving both modes 
(along and perpendicular to the brim) a nonvanishing curvature.
The hat  has lost its $U_A(1)$ symmetry along the vertical axis, but 
the original symmetry still remains in the sense that any rotated state
of the hat along the vertical axis is an equally probable final configuration.
But this degree of freedom does not correspond to the meson  masses.}
in such a way that the axial vector 
symmetry in the original tree-level Lagrangian  is broken. One obtains
when including quantum effects
\be
 {\cal L'} = {\cal L} + \frac 1 2 \Delta m_0^2 s_0^2  \ ,
\ee
where ${\cal L}$ now does not include ${\cal L}^{SB}$. Instead,  the 
new term is generated through the loops of Fig.~5. It looks just like
a term which explicitly breaks the symmetry, but is in fact,  generated
through the self-consistency condition for the potential. 
(Formally one could eliminate the new term by  adding, by hand, 
a renormalization counter term adjusted in such a way that the new term is 
exactly cancelled. But then one would have to again add a similar term, 
by hand, in order to give the pseudoscalars mass. 
Such a procedure would of course be ridiculous; 
it is more natural to consider the tree level Lagrangian (1)
to be fundamental, but its symmetries broken by quantum effects.)  

Clearly Ward
identities involving the divergence of the axial vector current will
look different when derived from ${\cal L} $ than from ${\cal L}'$.
With the new term in ${\cal L}'$\  the Ward identities for the
divergence of the axial vector currents  will look just like the 
conventional ones where a pseudoscalar mass  (or
quark mass in QCD) is put in by hand. 
The main difference is that now the symmetry violating term is not put in by
hand, but is evaluated through the self-consistency 
condition from the three level  Lagrangian (1). 
Assuming the usual relations between
quark masses  and pseudoscalar masses in QCD,
this would imply spontaneous generation of quark masses.
This mechanism also opens up the door to a better understanding of 
the next step of symmerty breaking:
the spontaneous  breaking of $SU(3)$ 
flavour symmetry discussed in  previous papers \cite{NAT3}.

{\it 6. Spontaneous or explicit symmetry breaking. Goldstone bosons.}
Above I have called  the second step in the symmetry breaking spontaneous,
since it follows naturally from the first step once quantum effects are 
included into the classical Lagrangian (1). However, in the terminology of
t'Hooft\cite{hooft}  symmetry breaking should be called spontaneous only if 
there appears Goldstone bosons, otherwise the symmetry breaking is explicit.
Therefore t'Hooft calls the mass generation of the $\eta'$ through 
the quantum effects related to the gluon anomaly an explicit symmetry breaking.
If one adopts this convention also our symmetry breaking should be called 
explicit, i.e. not spontaneous, since also in our case the symmetry breaking
is quantum mechanical and no related  Goldstone bosons  seem to exist.

We know experimentally that no additional scalar
massless bosons related to our second step of chiral symmetry breaking
have been observed (at least not as free particles, i.e. not counting confined 
ghosts, schizons or gluons). How then,
can the Goldstone theorem  be circumvented? We note that the
simplest way out is to observe, that in Eq.~(1) one really  does
not need the full {\it continuous} U3$\times $U3 symmetry. 
One can replace it with a {\it discrete} permutation symmetry where parity 
and flavour indices are permuted  and still get the same constraints with
degenerate mass for the whole scalar and pseudoscalar nonets before any symmetry breaking takes place. It is true that
it is practical to embed this  permutation symmetry into a larger continuous
unitary group, but it is really not necessary.
 
Furthermore, one should remember that we do have superselection rules\cite{ww}
of parity, charge, and generally flavour in strong interactions. 
E.g. superpositions of  different charge states, say $\pi^+$
and  $\pi^0$, or superpositions of 
$\pi^+$ and $a_0^+$, are  not  physically realizable states, 
in the same way as, say, different 
spin states are. I.e., the physical Hilbert space only includes the discrete
states of definite flavour and parity, 
not superpositions of these, which are generated by the full continuous group.
With this limitation in mind it is, in fact, more natural to look at 
flavour and chiral symmetry of strong interactions 
as a discrete symmetry. Then there is no conflict
with the Goldstone theorem, since a discrete symmetry can broken 
spontaneously (or explicitly) without the appearance of Goldstones.

{\it 7. Concluding remarks.}
Finally my approach has two extra benefits which I find worth mentioning: 
(i) it puts the "mysterious"
OZI rule and its breaking through loops on a firmer Lagrangian framework
through the dominance of the $\lambda$ term at the tree level in Eq.~(1), 
and (ii) it may provide a resolution to the strong CP problem,
since a quark mass can be put zero in the original tree level Lagranian, 
although  an effective mass is generated through the spontaneous chiral
symmetry breakdown in loops.

{\it Acknowledgements.} I thank C. Montonen, D.O. Riska, M. Roos and M. Sainio
for useful comments, and H. Leutwyler, E. de Rafael and S. Weinberg for e-mail
correspondence emphasizing the problem of Goldstone bosons, 
which led to the inclusion of section 6.

\end{document}